\documentclass[aps,prl,twocolumn,amsmath,amssymb,showpacs,superscriptaddress]{revtex4}
\usepackage[dvips]{graphicx,psfrag}
\usepackage[usenames]{color}
\usepackage[english]{babel}
\usepackage{latexsym}
\usepackage{subfigure}
\usepackage{epsfig}
\usepackage{epstopdf}
\usepackage{ulem}
\begin{document}

\title{Compression as a tool to detect Bose glass in cold atoms experiments}

\author{Dominique Delande}

\affiliation{Laboratoire Kastler-Brossel, Universit\'e Pierre et Marie Curie-Paris 6, ENS, CNRS;  
4 Place Jussieu, F-75005 Paris, France}

\author{Jakub Zakrzewski} 

\affiliation{Laboratoire Kastler-Brossel, Universit\'e Pierre et Marie Curie-Paris 6, ENS, CNRS;  
4 Place Jussieu, F-75005 Paris, France}

\affiliation{
Instytut Fizyki imienia Mariana Smoluchowskiego and
Mark Kac Complex Systems Research Center, 
Uniwersytet Jagiello\'nski, ulica Reymonta 4, PL-30-059 Krak\'ow, Poland}

\date{\today}

\pacs{03.75.Lm,67.85.Hj,03.75.Kk}

\begin{abstract}
We suggest that measuring the variation of the radius of an atomic cloud
when the harmonic trap confinement is varied makes it possible to monitor the 
disappearance of the insulating Mott phase of an ultracold
atomic gas trapped in a disordered optical lattice. This paves
the way for an unambiguous identification of a Bose glass phase in the system.
\end{abstract}
\maketitle

Condensed matter models have found recently a wonderful testbed in
cold atoms in optical lattices physics~\cite{jaksch05,lewen2007}. Cold atoms
allow for an optimal control of parameters of the system:  by changing 
intensities or detunings of laser beams, one may modify  the depth of
the optical lattices; interactions between atoms can be controlled
via external magnetic 
field and Feshbach resonances. An example
is the Bose-Hubbard tight binding model \cite{fisher89} which can be realized
using ultracold atoms and where the quantum phase transition between a superfluid (SF) 
phase and a Mott insulator (MI) phase 
can be observed by varying the depth of the optical lattice
as predicted in a seminal work \cite{jaksch98} and realized few years later \cite{greiner02}. 

Controllability of the system makes studies of disorder induced effects particularly interesting.
Disorder (or pseudo-disorder) may be created in cold atoms systems in a repeatable way
using optical potentials created by laser speckles or multichromatic lattices \cite{damski03,roth03}.
This exciting possibility attracted soon a lot of research \cite{paper-speckles} - Anderson
localization being one of the main targets for weakly or non-interacting atoms~\cite{aspect08,ingu08}. 
Even more exciting is another regime of strongly interacting bosons in a disordered potential.  
Seminal studies of the disordered Bose-Hubbard model \cite{giamarchi88,fisher89} revealed the 
existence of a novel insulating phase called Bose glass (BG) phase. Contrary to the MI, the BG 
is characterized by gapless excitation spectrum and is compressible. As far as we know it has
not been observed in a ``traditional'' condensed matter settings. Its unambiguous observation
is certainly an important milestone yet to be seen.
  
In the first attempt to produce a BG with ultracold atoms, a bichromatic 
quasi-disordered  optical lattice was used \cite{fallani07}. 
The authors measured both the absorption spectrum of the system and the long-range 
spatial coherence.
They observed a smearing of the absorption peaks in the presence of ``disorder''
as well as a decreased long-range coherence, which they interpreted as a manifestation of the existence of a BG phase. 
The interpretation of absorption spectra is however difficult: 
as discussed by us elsewhere \cite{zakdd08}, the initial state was not the ground state of the system
(this would be the case if the lattice were ramped up adiabatically as supposed in~Ref.~\cite{fallani07}).
Also
the amplitude of the lattice modulation was strong, so that the absorption was not in the linear regime. 

Another very recent attempt \cite{white08} has shown -- in the three-dimensional case -- that the presence of the disorder 
leads to a significant decrease of the condensate fraction both for SF or coexisting SF and MI phases.
This measurement follows in fact the original proposition of \cite{damski03} to address the
disappearance of the condensate fraction as a possible signature of the BG presence. Here,
again the preparation of the initial state is a key question. 

While both these experiments are important studies of strongly
interacting bosonic systems in the presence of disorder, it is desirable to have a clear
signature of a BG phase. 
Of special interest is a direct measurement of the compressible
or incompressible nature of the system.
An interesting possibility is to access the central density 
of the atomic cloud and measure the changes of that density when the external trapping potential is varied
(trap squeezing spectroscopy) \cite{roscilde08}. That measurement supplemented with coherent fraction 
measurement provides an interesting and clear proposition for an ``ideal'' experimental procedure.
There are two requirements. One is to make the trap geometry independent from laser beams forming the lattice, 
so that one can vary the trap frequency independently of the lattice
height. The second one is to use an additional focused laser to monitor
the density in the center of the trap. The latter while feasible seems quite difficult. 
The aim of this letter is to propose a much simpler 
experimental scenario, which has already been used in demonstrating the Mott phase for
a cloud of fermions \cite{schneider08}. We propose to measure the radius of the atomic cloud while changing the trap frequency.
This provides a simple and direct measurement of the compressibility of the system! More precisely, it makes it
possible to monitor the appearance and disappearance of incompressible phases. Complemented with 
measurements of the long-range phase coherence, it would allow for an unambiguous characterization of
the MI-BG-SF phase diagram.

This method works beautifully for fermions \cite{schneider08} where in the Mott state one has at most
one fermion per spin state. In such a case the radius of the cloud becomes practically independent of the 
trap frequency, clearly demonstrating the incompressibility of the MI state.
The situation is quite different for bosons where in typical
experiments \cite{greiner02,fallani07} one may have up to three bosons per site of the optical lattice and
the density profile in the trap resembles that of a wedding cake - see e.g. \cite{Batrouni02}. 
Between Mott regions
with  single, double, or triple  occupancies, there are superfluid ``shells''. Those  
lead to a finite compressibility of the total sample and could make radius measurements useless. 
As shown below, this is not a major problem.

Let us start with a one-dimensional Bose-Hubbard model in the presence of a trapping potential
and a diagonal disorder. The Hamiltonian is~\cite{jaksch98}:
\begin{equation}
\hat{H} = -J \sum_{\left<j,j'\right>}
\hat{b}_j^\dagger \hat{b}_{j'} + \frac{U}{2} \sum_{j} \hat{n}_j
\left( \hat{n}_j - 1 \right) + \sum_{j} \epsilon_j \hat{n}_j, 
\label{hamiltonian}
\end{equation}
where $\hat{b}_j$ ($\hat{b}_j^\dagger$) is the destruction (creation) operator of one 
particle in the $j$-th site,
$\hat{n}_j=\hat{b}_j^\dagger \hat{b}_j$ is the number operator, and $\left<j,j'\right>$ 
indicates the sum on nearest neighbors.
$U$ is the interaction energy and $J$ the hopping energy. 
The energies at sites, $\epsilon_j$ are given by the 
sum of the energy shift due to the the harmonic potential and the additional disorder:
\begin{equation}
\epsilon_j=\frac{1}{2} m \omega^2 a^2 (j-j_0)^2 + x_j U \Delta
\end{equation}
where $m$ is the particle mass, $a$ the lattice spacing, $\omega$ the trap frequency
and $j_0$ the position of the trap center. $\Delta$ is a dimensionless parameter measuring the strength of the disorder
(in units of the interaction energy), while $x_j$ is a (pseudo-)random variable.
We consider two different types of disorder.
For a truly random disorder, the $x_j$'s are chosen as independent variables
with uniform distribution in the interval [-1,1].
For a secondary optical lattice as used in~\cite{fallani07}, $x_j$ is
a sine function with incommensurate frequency: $x_j=\sin(\lambda j)$ resulting
in pseudo-random {\it correlated} variables with distribution~\cite{guarrera07}:
\begin{equation}
P(x)=\frac{1}{\pi \sqrt{1-x^2}}.  
\label{disorder_florence}
\end{equation}
The system properties in such a pseudo-random disorder may differ from the truly
random situation~\cite{kollath08,Roscilde07,Deng08}.

We employ the parameters originating from
the Florence \cite{fallani07} experiment with the  exception that we assume the possibility of independent change of trap frequencies. 
In particular we concentrate on the deep Mott regime ($J/U\approx 0.027$).

As shown below, in this regime -- and as long as we look at the compressibility of the system and not
at long range phase coherence -- we can use a local density approximation (LDA) to describe the system
~\cite{schneider08,guarrera07}.
This amounts at neglecting tunneling between neighboring sites and assuming a Fock state at each site with the occupation determined by the
local chemical potential $\mu-\epsilon_j.$
Determining the shape of the atomic cloud is a simple minimization
procedure for the total energy, i.e., the sum of independent contributions for various sites, constrained
by a fixed total number of particles.

In the absence of disorder and at low trapping frequency, the ground state is a pure MI with
an unit occupation number at all sites around the trap center. If we define the r.m.s radius
of the trap (in units of the lattice spacing $a$) as:
\begin{equation}
R=\sqrt{\langle r^2\rangle - \langle r \rangle^2} = \sqrt{\frac{\sum_j{j^2 n_j}}{N} - \left(\frac{\sum_j{j n_j}}{N}\right)^2}
\end{equation}
where $N$ is the total number of particles, it is clear that, for large $N$ and low trap frequency, the radius
will be $R_c=N/2\sqrt{3}$ {\it independently} of the trap frequency, a clear-cut manifestation
of the incompressibility of the MI phase.
When the trap frequency is increased, the energy of the outest particles $m\omega^2a^2N^2/8$ increases until the point where
it is cheaper to pay the interaction energy $U$ and transfer the particle at the trap center, creating
doubly occupied sites. Beyond this point, $R$ decreases with frequency,
implying global compressibility.
In the large $N$ limit, a straightforward calculation shows that the parameters obey scaling laws.
The critical frequency is (from the two estimates above) $\omega_c=\sqrt{\frac{8U}{ma^2}}\frac{1}{N}.$
We thus define the scaled frequency:
$\omega_0=\omega/\omega_c$
and the scaled r.m.s. radius,
$R_0 = R/R_c$.

\begin{figure}
\begin{center}
\psfrag{radius}{\large$R_0$}
\psfrag{omega0}{\large$\omega_0$}
\includegraphics[width=0.4\textwidth]{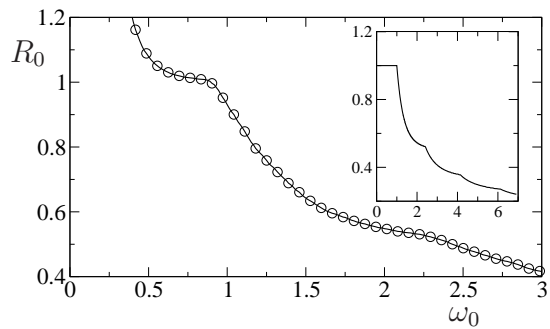}
\end{center}
\caption{Scaled r.m.s. radius of the one-dimensional atomic cloud as a function of the scaled trap 
frequency for a random uniform disorder with $\Delta$=0.2.
The kink around $\omega_0$=1 is a signature of the appearance of a new phase.
The solid line is obtained using the local density approximation (LDA). The circles are the results
of a quantum calculation with the TEBD algorithm for 61 particles and $J/U$=0.027. 
The inset shows the scaled r.m.s. radius on a wider scale, in the absence of disorder.
One clearly sees the additional kinks related to the appearance of  doubly and triply
occupied MI phases.}
\label{fig_1d_vidal}  
\end{figure}

A plot of $R_0$ versus $\omega_0$ is shown in the inset of Fig.~\ref{fig_1d_vidal}, for $N$=151 in the absence of disorder.
For shallow trap, the radius shows a pronounced plateau. In this range of frequencies, all particles are in the 
Mott phase with an unit filling - the plateau
is a direct manifestation of incompressibility. This regime resembles most the fermionic case
\cite{schneider08}. At $\omega_0$=1, a sharp kink indicates the moment when the sample
becomes compressible - at this point a double occupancy appears at the centre of the trap
(this can be visualized looking directly at
the occupation of sites). The compressibility becomes smaller for larger frequency reaching a second kink
at $\omega_0=1+\sqrt{2}\approx 2.414 $ - here a triple occupancy in the center of the trap is born. 
A careful inspection reveals even the third kink at $\omega_0=1+\sqrt{2}+\sqrt{3}\approx 4.146$.
From this plot, it is clear that measuring the {\it global} quantity $R$ vs. trap frequency makes it possible
to monitor the appearance of the successive MI phases. Let us note that for the Florence experiment
\cite{fallani07} the effective harmonic binding (coming from the trap and lasers' transverse profiles)
is 75~Hz, corresponding to $\omega_0\approx3.44$. 

In the presence of a random disorder, the plateau disappears, showing a non zero
compressibility of the sample, as seen in the main panel of Fig.~\ref{fig_1d_vidal}. 
At low frequency, this is due to the appearance of a compressible
BG phase in the external part of the cloud (where the average
occupation number is between 0 and 1). The kink around $\omega_0$=1 is, however, a robust feature. 
It is due
to the formation of first a BG with occupation number between 1 and 2 at the trap center,
followed by the birth of a $n$=2 MI phase. The kink itself is the signature of the appearance
of new phases. 
The residual  tunneling between the neighbouring sites has almost no effect. To prove this statement, 
we use the TEBD algorithm \cite{vidal03}  (also known as t-DMRG algorithm \cite{tdmrg}) 
 with imaginary time propagation to produce
the quasi-exact ground state of the Bose-Hubbard Hamiltonian in the presence of disorder
from which the corresponding r.m.s. radius easily computed.  
The result, averaged over several realizations of the disorder, is shown in Fig.~\ref{fig_1d_vidal}.
It is almost indistinguishible from the result of the LDA approximation. We have checked that, for other
disorder strengths -- but still in the low tunneling regime -- there is a similar agreement between the 
exact
TEBD result and the LDA approximation, which we will use in the following of this paper.
Most probably, this is because tunneling creates coherence between neighbouring sites and 
smoothes the steps of the wedding cake,
but does not induce macroscopic transfer of particles over long distances and thus only
marginally affects the r.m.s. radius. 
From the point of view of compressibility, whether a BG or a SF phase
is formed makes little difference.

\begin{figure}
\begin{center}
\psfrag{R0}{\large $R_0$}
\psfrag{omega0}{\large $\omega_0$}
\includegraphics[width=0.4\textwidth]{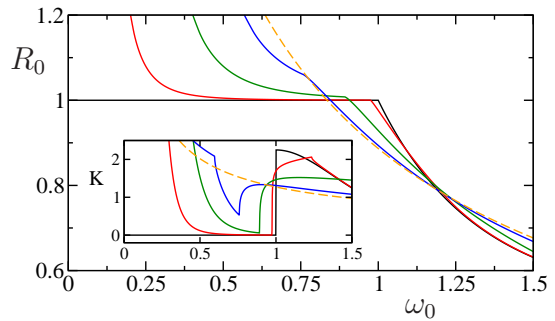}
\end{center}
\caption{(Color online) Scaled r.m.s. radius of the one-dimensional atomic cloud as a 
function of the scaled trap frequency for a random uniform disorder.
The successive curves (from left to right) are for an increasing disorder strength
$\Delta=0, 0.05, 0.2, 0.4$ and 0.5. The kink in the curves disappears for the limiting case 
of $\Delta$=0.5 (dashed line) beyond which the MI phase no longer exists. 
The inset shows the dimensionless global compressibility, eq.~(\ref{def_K}), which displays
a marked minimum related to the existence of the incompressible MI phase.}
\label{fig_1d}  
\end{figure}

The MI phase is expected to disappear at $\Delta$=0.5 in the limit of small tunneling \cite{fisher89}.
In Fig.~\ref{fig_1d}, we show $R_0$ vs. $\omega_0$ for increasing disorder strengths. The kink is very clearly 
visible even at values very close to $\Delta$=0.5, showing that the proposed simple  method allows one
to monitor appearance and disappearance of various phases.
At the critical disorder $\Delta$=0.5, the MI phases completely disappears and a smooth curve is obtained.
It can be shown~\cite{long_paper} that its equation is: $R_0=2^{-1/3}3^{5/6}5^{-1/2}\omega_0^{-2/3}.$

The kink is even more visible if one considers the compressibility of the system, that is the derivative
of the radius with trap frequency. In ~\cite{schneider08}, a global compressibility
$\kappa_R = -\frac{1}{R^3} \frac{\partial R}{\partial (m\omega^2a^2/2)}$ is defined, 
where the powers of $R$ 
are chosen to obtain well-defined quantities in the thermodynamic limit $N\to \infty.$ As it has the
dimension of the inverse of an energy we prefer to use a dimensionless
global compressibility $K$ by multiplying $\kappa_R$ by $U.$ It has a simple expression in terms of scaled quantities:
\begin{equation}
K = - \frac{U}{R^3} \frac{\partial R}{\partial (m\omega^2a^2/2)} = - \frac{3}{2R_0^3\omega_0} \frac{dR_0}{d\omega_0}
\label{def_K}
\end{equation}

It is shown in the inset of Fig.~\ref{fig_1d} for various disorder strengths. While it vanishes below $\omega_0$=1
for zero disorder, it displays a well marked minimum for $0<\Delta<0.5,$ directly related to the
existence of an incompressible MI phase. At $\Delta=0.5,$ the MI phase and the minimum disappear.
Note that the observed behaviour is quite similar to the one observed for fermions in~\cite{schneider08},
the role of tunneling being here replaced by disorder.

\begin{figure}
\begin{center}
\psfrag{R0}{\large$R_0$}
\psfrag{omega0}{\large$\omega_0$}
\includegraphics[width=0.4\textwidth]{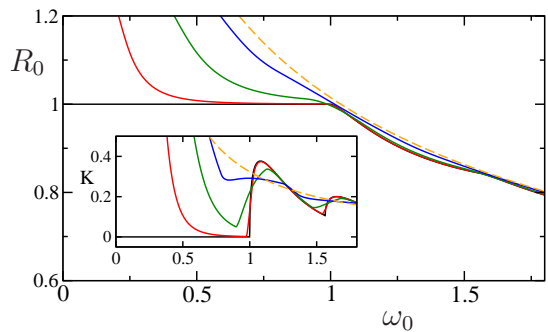}
\end{center}
\caption{(Color online) Same as Fig.~\ref{fig_1d}, for the three-dimensional disordered isotropic Bose-Hubbard model. The behaviour
is similar, with a kink in the curve around $\omega_0$=1 and a marked minimum of the compressibility (inset).}
\label{fig_3d}  
\end{figure}

A similar behaviour is observed for the three dimensional isotropic Bose-Hubbard
model. The results are shown in Fig.~\ref{fig_3d}. The singularity is slightly different
(the compressibility has no discontinuity, but a very fast increase), but the essential property:
the existence of a kink in the $R_0$ vs. $\omega_0$ curve, or equivalently, the marked minimum
in the compressibility, is present as well. The scaling laws are of course
different in three dimensions: we now have $R_c=3^{5/6}5^{-1/2}2^{-1/3}\pi^{-1/3}N^{1/3}$
and $\omega_c=2^{7/6}\pi^{1/3}3^{-1/3}N^{-1/3} \sqrt{U/ma^2}$ and the 3/2 coefficient
in eq.~(\ref{def_K}) must be replaced by 5/6. The appearance of the second MI phase
is clearly  visible as a dip in the compressibility around $\omega_0=1.5$.

\begin{figure}
\begin{center}
\psfrag{R0}{\large$R_0$}
\psfrag{omega0}{\large$\omega_0$}
\includegraphics[width=0.4\textwidth]{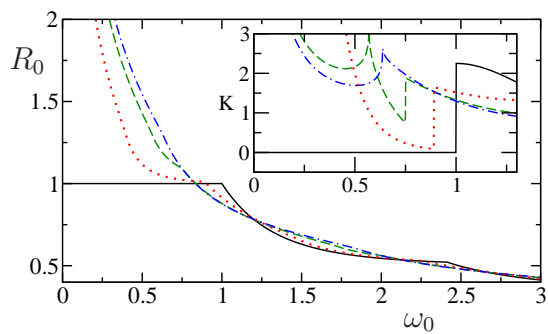}
\end{center}
\caption{(Color online) Same as Fig.~\ref{fig_1d} but for the pseudo-random ``disorder'' created by a secondary
incommensurate lattice: $\Delta$=0 (solid line), $\Delta$=0.2 (dotted line), $\Delta$=0.4 (dashed line),
$\Delta$=0.5 (dash-dotted line beyond which the kink and the MI phase disappear). 
}
\label{fig_1d_florence}  
\end{figure}

Consider now the pseudo-disorder, eq.~(\ref{disorder_florence}), employed in the Florence experiment \cite{fallani07}. 
We have performed similar calculations,
averaging over the relative phase of the two lattices. 
An overall
similar behaviour, shown in Fig.~\ref{fig_1d_florence}, is observed, a pronounced kink yields the frequency when the double occupancy
emerges at the centre of the trap. With increasing disorder, the plateau shrinks, becomes
tilted, the kink moves towards smaller frequencies and eventually disappears at $\Delta$=0.5.
This translates in the second lattice depth equal to $s_2\approx 1.014$ (in the notations of \cite{fallani07}), which  nicely matches the estimate for the disappearance of the Mott phase and
the gap in the absorption spectrum \cite{fallani07}. 

Note that another singularity (sharp peak in the compressibility) is visible at lower frequency
and large disorder. This is due to the peculiarity of the distribution, eq.~(\ref{disorder_florence}),
diverging at $x=\pm 1.$  Those, however, may be 
easily identified and distinguished from the ``main'' Mott-end kink.

It has been argued that the MI regions may be strongly affected by temperature effects \cite{gerbier07}.
Our calculations are limited to $T=0$. Study of the effects of the temperature as well as of large tunneling on the
phenomenon discussed here are in progress.

In summary, we have shown that a simple measurement of the radius of the atomic cloud may provide a clear 
identification of the disappearance of the Mott phase and, in the presence of disorder, 
may help to unravel the BG presence. The disappearance of the MI is correlated with vanishing of the 
kink in the radius-frequency dependence. DD acknowledges support by IFRAF, 
JZ acknowledges support by Polish Ministry of Education and Sports (2008-2011).
This work is realized within Marie Curie
TOK scheme COCOS (MTKD-CT-2004-517186).

\end{document}